\documentclass[11pt,a4paper,english,nofootinbib,,superscriptaddress]{revtex4}
\usepackage{lmodern}

\usepackage[T1]{fontenc}
\usepackage[latin9]{inputenc}
\usepackage{babel}
\usepackage{color}
\usepackage{cancel}
\usepackage{amsmath}
\usepackage{graphicx}
\usepackage{amssymb}
\usepackage{esint}
\usepackage[unicode=true, pdfusetitle,
bookmarks=true,bookmarksnumbered=false,bookmarksopen=false,
breaklinks=false,pdfborder={0 0 1},backref=false,colorlinks=false]
{hyperref}
\usepackage{amsmath}
\usepackage{amssymb}
\usepackage[compat=1.0.0]{tikz-feynman}
\usepackage{float}

\def\b{\begin{equation}} \def\e{\end{equation}}
\def\bd{\begin{displaystyle}} \def\ed{\end{displaystyle}}
\def\ba{\begin{array}} \def\ea{\end{array}}

\def\bee{\begin{enumerate}}
	\def\eee{\end{enumerate}}

\def\ud{\mathrm{d}}

\def\1{\mbox{I\hspace{-.15em}1}}

\def\R{{\rm I\hspace{-.15em}R}}

\def\b{\begin{equation}}
\def\e{\end{equation}}
\def\bee{\begin{enumerate}}
	\def\eee{\end{enumerate}}

\makeatletter
\usepackage{latexsym}\usepackage{bm}

\makeatother

\begin{document}
	\title{Electron-photon interaction in \\ de Sitter ambient space formalism}
	
		\author{F. Jalilifard}
	\email{jalilifard@yahoo.com}\affiliation{Department of Physics, Razi University, Kermanshah, Iran}
	
	\author{Y. Ahmadi}
	\email{ahmadi.pave@gmail.com} \affiliation{Department of
		Physics, Razi University, Kermanshah, Iran}

		\author{M.V. Takook}
	\email{takook@razi.ac.ir} \affiliation{Department of Physics, Razi University, Kermanshah, Iran}
	
\begin{abstract}
The electron-photon interaction (the spinor-vector field interaction) in the de Sitter ambient space formalism is investigated in the first (leading order) approximation. The interaction Lagrangian and the scattering matrix are presented. In this approximation, the scattering matrix can be written as a expansion of the interaction Lagrangian. The tree-diagram of electron-photon scattering amplitude is calculated. Finally the Minkowski limit is considered.
\end{abstract}
	
\maketitle
\section{Introduction}
The effect of the gravitational field on the quantum fields is an interesting problem for understanding the behavior of the quantum gravity in the linear approximation. In this direction, the quantum field theory (QFT) in curved space-time and  the quantum black hole permit us to better understand the quantum gravity effect. Therefore it is very important that the QFT in de Sitter (dS) space-time is studied. The cosmic microwave background radiation \cite{Nature}, highly redshift observation of the Supernova Ia \cite{Perl, Riess} and galaxy clusters \cite{Henry 1, Henry 2}, indicate that our current universe in the first approximation may be described by a dS space-time. The observational data  collected by BICEP2 \cite{BICEP2} may confirm that the early universe, in a good approximation, is also the dS universe. Therefore, in this paper the background curved space-time is chosen as the dS space-time.

The dS space-time can be considered as 4-dimensional hyperboloid embedded in 5-dimensional flat Minkowsi space-time, which is called the ambient space formalism. Because of linearity of the action of dS group on the ambient space formalism, unlike the intrinsic coordinates, working with this formalism leads to simplicity of formulation and calculation of the Green functions. The unitary irreducible representation of dS group and the analyticity of the complexified dS space-time is a base of the rigorous mathematical construction of QFT in dS space-time that have been studied in \cite{77,Bros 1,brmo,brmo03,tak,ta97}.

The unitary irreducible representation is used to consider the free field and to construct the Hilbert space and Fock space. In Minkowski flat space-time, the analyticity properties are the basis parts of the interaction fields calculations, and the theory of interacting fields is formulated in terms of global analyticity properties of the two-point functions. The analyticity of the two-point function is a very important properties which permit us to calculate the probability amplitude in the interaction case.
 These properties for dS space-time was proved by Brose \cite{Bros 1,brmo,brmo03} and in dS ambient space formalism these are very similar to the Minkowski space-time. Thus it is suitable to use this formalism to consider the interaction between the different fields. 

 In this paper, the quantum electrodynamics in dS universe is discussed. In this direction the electron-photon interaction (the spinor-vector fields interaction) is studied. After fixing our notations in section \ref{notations}, which include a description of the free spinor and vector field quantization in the dS ambient space formalism, the interaction Lagrangian and the scattering matrix are presented in section \ref{interaction}. Section \ref{compton} is devoted to the Compton scattering. Finally, the Minkowski lime is presented in section \ref{mili}. We discuss our results in section \ref{conclu}.

\setcounter{equation}{0}
\section{Notations and terminology}  \label{notations}
	
The dS space-time can be considered as a four-dimensional hyperboloid embedded in five-dimensional Minkowski space-time with the following equation:
	$$
	M_H=\left\{ x \in \R^5| \; \; x \cdot x=\eta_{\alpha\beta} x^\alpha
	x^\beta =-H^{-2}\right\},\;\; \alpha,\beta=0,1,2,3,4\,.$$
The metric is:
	\b \label{ds difiniion}
	ds^2=\eta_{\alpha\beta}dx^{\alpha}dx^{\beta}|_{x^2=-H^{-2}}=g_{\mu\nu}^{dS}dX^{\mu}dX^{\nu},\;\; \mu=0,1,2,3,\e
	where $\eta_{\alpha\beta}=$diag$(1,-1,-1,-1,-1)$, $H$ is Hubble parameter, $X^\mu$ is dS intrinsic coordinates and $x^\alpha$ is the five-dimensional dS ambient space formalism.
	
	The dS group is defined by $
	SO(1,4)=\left\lbrace \Lambda \in GL(5,\R)| \;\; \det \Lambda=1,\;\; \Lambda \eta \lambda^t= \eta \right\rbrace,$
	where $\Lambda^t$ is the transpose of $\Lambda$.
	The action of dS group on the ambient space coordinate $x^\alpha$ is linear and this leads to simplicity of equation and calculation in this formalism \cite{77}:
	\b \label{II.4}
	x'^\alpha=\Lambda^\alpha_{\;\;\beta} x^\beta, \;\;\;\Lambda \in SO(1,4) \Longrightarrow x\cdot x=x'\cdot x'=-H^{-2}. \e
In this formalism the five matrices $\gamma^{\alpha}$ are needed which satisfy the following relation \cite{ta97,ta96,bagamota,77}:
	\b \label{II.8}
	\gamma^{\alpha}\gamma^{\beta}+\gamma^{\beta}\gamma^{\alpha}
	=2\eta^{\alpha\beta}\;,\;
	\gamma^{\alpha\dagger}=\gamma^{0}\gamma^{\alpha}\gamma^{0}\, ,\e
	and they may be chosen as:
	$$
	\gamma^0=\left( \begin{array}{clcr} I & \;\;0 \\ 0 &-I \\ \end{array} \right)
	,\;\;\;\gamma^4=\left( \begin{array}{clcr} 0 & I \\ -I &0 \\ \end{array} \right),\;$$ \b \label{gamma relation} \gamma^1=\left( \begin{array}{clcr} 0 & i\sigma^1 \\ i\sigma^1 &0 \\
	\end{array} \right)
	,\;\;\gamma^2=\left( \begin{array}{clcr} 0 & -i\sigma^2 \\ -i\sigma^2 &0 \\
	\end{array} \right)
	, \;\;\gamma^3=\left( \begin{array}{clcr} 0 & i\sigma^3 \\ i\sigma^3 &0 \\
	\end{array} \right)\, ,\e
	where $\sigma^i$ $(i=1,2,3)$ are the Pauli matrices. The $\gamma^\alpha$ matrices in dS ambient space formalism are different from Minkowski $\gamma'^\mu$ matrices. In the null curvature limit the relation between them are \cite{bagamota}:
	\b \label{minkofski gamma}\gamma'^{\mu}=\gamma^{\mu}\gamma^4.\e
 
The dS Dirac first-order field equation is \cite{77,ta97}:
	\b \label{II.34}
	\left(\cancel{x} \;\cancel{\partial}^\top-2 \pm i\nu\right)\psi(x)=0,\;\;\; \cancel{x}=\gamma_\alpha x^\alpha=\gamma\cdot x\,,\e
	where $\nu>0$ is related with dS mass parameter as $m_{f,\nu}^2=H^2(\nu^2+2\pm i\nu)$ and $\partial_{\alpha}^\top=\partial_{\alpha}+Hx_{\alpha}x.\partial$ is the transverse derivative.
	The charged spinor field operator, which satisfy the above field equation, is \cite{bagamota,77}:
\b \label{psi expansion}
	\psi(x)=\int_{S^3}d\mu(\xi)\sum_{\sigma}\left[a( \tilde{\xi},\sigma)(Hx.\xi)^{-2-i\nu} {\cal U} (x,\xi,\sigma)+b^{\dag}(\xi,\sigma)(Hx.\xi)^{-1+i\nu} {\cal V} (x,\xi,\sigma)\right],\e
	where $\xi^\alpha=(\xi^0, \vec \xi, \xi^4)\in C^+=\left\lbrace \xi \in \R^5|\;\; \xi\cdot \xi=0,\;\; \xi^{0}>0 \right\rbrace$  is the transformed variable of $x^{\alpha}$ in positive cone that correspond to energy-momentum four-vector in Minkowski space-time. $\tilde{\xi}^\alpha=(\xi^0, -\vec \xi, \xi^4)$ and the $\ud \mu(\xi)$ is the $SO(4)$-invariant normalized volume. The explicit form of $\cal{U}$ and $\cal{V}$ were presented in \cite{bagamota}. The adjoint spinor $\overline{\psi}(x)$ in ambient space formalism is define as $\overline{\psi}(x)=\psi^\dagger(x)\gamma^0\gamma^4$ \cite{ta96,bagamota,ta97,77}.
	The creation operators can be defined as \cite{77}:
		\b \label{II.27}
	a^\dag (\xi,\sigma) \left| \Omega \right> \equiv\left|1_{\xi, \sigma}^{a} \right\rangle,\;\; b^\dag (\xi,\sigma) \left| \Omega \right> \equiv\left|1_{\xi, \sigma}^{b} \right\rangle\, .\e
	The vacuum state $|\Omega>$ is invariant under the action of the UIR of the dS group and it can be identified with the Bunch-Davies or Hawking-Ellis vacuum state. Its norm can be fixed in the null curvature limit as $\left< \Omega \right.\left| \Omega \right>=1 $ \cite{brmo,77}. The anti-commutation relations for creation and annihilation operators are:
	$$\left[a(\tilde{\xi}',\sigma') , a^\dagger(\xi,\sigma)\right]_{+}={\cal N}_p(\xi,\sigma)\delta_{s^3}(\xi-\xi')\delta_{\sigma \sigma'}\, ,$$
	\b \label{anti-commutation relation}
	\left[b(\tilde{\xi}',\sigma') , b^\dagger(\xi,\sigma)\right]_{+}={\cal N}_p(\xi,\sigma)\delta_{s^3}(\xi-\xi')\delta_{\sigma \sigma'}, \e
	where ${\cal N}_p(\xi,\sigma)$ is the normalization constant.
	
	Analytic field operator is define in complex dS space-time as \cite{77,brmo}:
$$\psi(x)=\lim_{y\rightarrow 0} \psi(z)=\lim_{y\rightarrow 0} \psi(x+iy). $$
The analytic two-point function of spinor field is:
	\b \label{spinor two-point function}	\left<\Omega\right|\psi_{i}(z_1)\bar \psi_{j}(z_2)\left|\Omega\right>=iS_{ij}(z_1,z_2)=\dfrac{c_{\frac{1}{2},\nu}}{2}\int_{S^3} d\mu(\xi) (z_1.\xi)^{-2-i\nu}(z_2.\xi)^{-2+i\nu}\left(\cancel{\xi}\gamma^4\right)_{ij},\e
	where 
	$c_{\frac{1}{2},\nu} =\frac{\nu(\nu^2+1)}{(2\pi)^3(e^{2\pi \nu}-1)}\, .$
	The analytic two-point function has been calculated in terms of generalized Legendre function of first kind \cite{ta96,ta97,bagamota}.
	
	In the ambient space formalism one can write the massless vector field, similar to other massless quantum field, in terms of the massless conformally coupled scalar field. The field operator obtain in terms of annihilation $d$ and creation $d^\dagger$ operators as \cite{gagarota,ta97}:
	\b \label{vector field expressed} A_\alpha(x)=\int_{B}d\mu({\xi})\sum_{n} \left[d({\tilde{\xi}},n)(Hx.\xi)^{-2}{\cal E}_{2\alpha}(x,\xi,n)+d^{\dag}(\xi,n)(Hx.\xi)^{-1}{\cal E}_{1\alpha}(x,\xi,n)\right]\, ,\e
	where $n=0,1,2,3$ are the polarization states and $d\mu({\xi})_B=2\pi^2r^3drd\mu({\xi})$ is the Euclidean measure on the unit ball $B$ \cite{tak,ta97}. The commutation relation between creation and annihilation operators is \cite{gata,gagarota}:
		\b \label{commutation relation}
	\left[d(\tilde{\xi}',n') , d^\dagger(\xi,n)\right]_{-}={\cal N}_k(\xi,n)\delta_{s^3}(\xi-\xi')\delta_{nn'}\, ,\e
	and ${\cal N}_k(\xi,n)$ is also the normalization constant.

	
\section {Scattering matrix} \label{interaction}
 
 In dS space-time, the dS-Dirac field equation is invariant under $U(1)$ global symmetry:
$$ \psi(x) \rightarrow \psi'(x)=e^{-i\epsilon}\psi(x),\;\;\;(-i\cancel{x}\; \cancel{\partial}^{\top}+2i+\nu)\psi'(x)=0\, ,$$
where $\epsilon$ is a constant. This equation is not invariant under the local $U(1)$ symmetry:
$$ \psi(x) \rightarrow \psi'(x)=e^{-i\epsilon(x)}\psi(x),\;\;\; (-i\cancel{x}\; \cancel{\partial}^{\top}+2i+\nu)\psi'(x)\neq0. $$ 
By changing the covariant derivative $\partial^\top_\alpha$ with a gauge covariant derivative $D_\alpha$:
$$ D_\alpha=\partial^\top_\alpha+iqA_\alpha ,\;\; \;\; A_\alpha\rightarrow A'_\alpha=A_\alpha+\dfrac{1}{Q}\partial^\top_\alpha\epsilon(x).$$
one can obtain dS-Dirac local gauge invariant equation \cite{agt2005}. 
In the free field Lagrangian by applying this changing, the interaction Lagrangian can be defined as \cite{agt2005}:
$$ {\cal L}_0 =H\bar{\psi}(x)\gamma^4(-i\cancel{x}\; \cancel{\partial}^{\top}+2i+\nu)\psi(x),\;\;\; {\cal L}_{int} =QH\bar{\psi}(x)\gamma^4 \cancel{x}\cancel{A}(x)\psi (x).$$
The $Q$ in the null curvature limit can be considered as the electric charge $e$. As it is seen, the interaction Lagrangian is independent of derivative of fields, therefor one can show easily that:
\b \label{vector spinor lagrangian} {\cal H}_{int}=-{\cal L}_{int}=-QH\bar{\psi}(x)\gamma^4 \cancel{x}\cancel{A}(x)\psi (x)\, .\e

The interaction Lagrangian has the important role in the theoretical and particle physics in Minkowski space-time for calculation of the probability amplitude or the S-matrix elements, but in curved space-time one can not define the asymptotic states ($t=-\infty$ and $t=+\infty$) for defining the S-matrix elements. For defining the time evolution operator, $\vert \alpha , t\rangle=U(t,t_0) \vert \alpha,t_0 \rangle$, at first, the time or coordinate system must be defined. For simplicity one can choose the static coordinate system:
 \b \left\{\begin{array}{clcr} x^0&=\sqrt{H^{-2}-r^2}\sinh Ht_s, \\                    
                      x^1&=\sqrt{H^{-2}-r^2}\cosh Ht_s, \\
                       x^2&=r\cos \theta , \\
                                  x^3&=r \sin \theta\cos\phi ,\\   
                                   x^4&=r\sin\theta\sin \phi,\\            
         \end{array} \right.\e
where $-\infty<t_s<\infty\;,\;0\leq r<H^{-1}\;,\;0\leq \theta\leq\pi\;,\;0\leq \phi< 2\pi$. This coordinate system does not cover all dS hyperboloid. In this coordinate system the metric is:      
         \b ds^2=\left(1-r^2H^2\right)dt_s^2-\left(1-r^2H^2\right)^{-1}dr^2- r^2 \left(d \theta^2 +\sin^2
              \theta d\phi^2\right).\e 

In this coordinate system, the time evolution operator for a quantum dS black-hole was considered (for more details see \cite{ta17}). Here since the interaction occur in the atomic dimension, we can use the flat limit approximation which ignore the direct effect of curvature in this dimension but the indirect effect exist. In this approximation the time evolution operator can be expanded in terms of the Minkowski counterpart:
$$U(t,t_0)={U_M}(t,t_0)+ Hf(t,t_0)+...\;,$$
where in the null curvature limit $U_M$ is exactly the time  evolution operator in Minkowski space-time. The term $f(t,t_0)$ is due to the direct effect of curvature which is ignored in this calculation since the interaction is in atomic level.

If we have the incoming free field state $|\phi_{in}\;,-\infty>$ which interact with the interaction Lagrangian ${\cal L}_{int}$, and the outgoing free field state $|\phi_{out}\;,+\infty>$, we can write $|\phi_{out}\;,+\infty>$ in terms of $|\phi_{in}\;,-\infty>$ as:
\b \label{smatrix} |\phi_{out}\;,+\infty>=U(\infty\;,-\infty)|\phi_{in}\;,-\infty>= {\cal S}|\phi_{in}\;,-\infty>.\e  
Therefore the scattering matrix may be presented in the following approximate equation to calculate the indirect effect of curvature corrections, from the Minkowski counterpart \cite{mandl shaw,kaku}:
$${\cal S}\simeq\frac{(-i)^{\lambda}}{\lambda!} \sum_{\lambda=0}^{\infty}{\cal S}^{(\lambda)} ,$$
where
\b \label{s matrix expansion} {\cal S}^{(\lambda)}=\int\ud\mu(x_1)\int\ud\mu(x_2)...\int\ud\mu(x_{\lambda})\; T[{\cal H}_{int}(\phi(x_1)){\cal H}_{int}(\phi(x_2))...{\cal H}_{int}(\phi(x_{\lambda}))]. \e
The symbol $T$ is the time order product of operators. The probability amplitude, Feynman amplitude, transition rate and other experimental quantity can be written by calculating the ${\cal S}$ matrix elements. It is important to note that in the null curvature limit the usual QED in Minkowski space-time is obtained.
 
 \section{Compton scattering}
\label{compton}
\begin{figure}[!ht]
\begin{tikzpicture}
\begin{feynman}
\vertex (aa);
\vertex [right=of aa] (ba){\(x_1\)};
\vertex [right=of ba] (ca);
\vertex [right=of ca] (ea){\(x_2\)};
\vertex [right=of ea] (fa);
\vertex [below=of aa] (f1a){\(\overline{\psi}^+\)};
\vertex [above=of aa] (p1a){\(A^-\)};
\vertex [below=of fa] (p2a){\(A^+\)};
\vertex [above=of fa] (f2a){\(\psi^+\)};

\vertex [right=of fa] (g);
\vertex [right=of g] (a);
\vertex [right=of a] (b){\(x_1\)};
\vertex [right=of b] (c);
\vertex [right=of c] (e){\(x_2\)};
\vertex [right=of e] (f);
\vertex [below=of a] (f1){\(\overline{\psi}^+\)};
\vertex [below=of c] (p1){\(A^+\)};
\vertex [above=of c] (p2){\(A^-\)};
\vertex [above=of f] (f2){\(\psi^+\)};

\tikzfeynmanset{
	every photon={red},
	every fermion={blue},
}

\diagram {
	(f1a) -- [fermion] (ba) -- [fermion, edge label'=\(S_F(x_1;x_2)\)] (ea)--[fermion](f2a),
	(p1a) -- [photon] (ba),
	(ea) -- [photon] (p2a),
};

\diagram {
	(f1) -- [fermion] (b) -- [fermion, edge label'=\(S_F(x_1 ; x_2)\)] (e)--[fermion](f2),
	(p1) -- [photon] (b),
	(e) -- [photon] (p2),
};

\end{feynman}
\end{tikzpicture}
\caption{Electron-photon scattering tree diagram}
\label{f}
\end{figure}
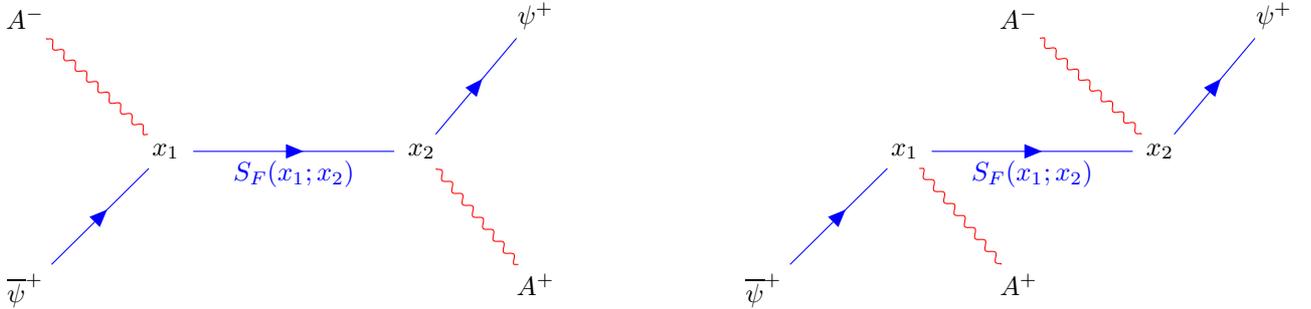

In this section we will study the Compton scattering or the interaction between the vector and spinor fields, in the tree level approximation. The scattering matrix ${\cal S}^{(2)}$ will be calculated for spinor and vector fields, which is very similar to the Minkowski QED.
By inserting the Lagrangian (\ref{vector spinor lagrangian}) in the equation (\ref{s matrix expansion}) for $\lambda=2$ and applying Wick's theorem \cite{mandl shaw}, the time order product of (\ref{s matrix expansion}) can be expressed in terms of the following normal order products:
\\
\\
$T\left[\overline{\psi}(x_1)\gamma^4\cancel{x}_1\cancel{A}(x_1)\psi(x_1) \overline{\psi}(x_2)\gamma^4\cancel{x}_2\cancel{A}(x_2)\psi(x_2)\right]=
N\left[\overline{\psi}(x_1)\gamma^4\cancel{x}_1\cancel{A}(x_1)\psi(x_1) \overline{\psi}(x_2)\gamma^4\cancel{x}_2\cancel{A}(x_2)\psi(x_2)\right]
\\
\\
+N\left[\underleftrightarrow{\overline{\psi}(x_1)\gamma^4\cancel{x}_1\cancel{A}(x_1)\psi(x_1) \overline{\psi}(x_2)\gamma^4\cancel{x}_2\cancel{A}(x_2)\psi}(x_2)\right]
+N\left[\overline{\psi}(x_1)\gamma^4\cancel{x}_1\cancel{A}(x_1)\underleftrightarrow{\psi(x_1) \overline{\psi}}(x_2)\gamma^4\cancel{x}_2\cancel{A}(x_2)\psi(x_2)\right]
\\
\\
+N\left[\overline{\psi}(x_1)\gamma^4\cancel{x}_1\underleftrightarrow{\cancel{A}(x_1)\psi(x_1) \overline{\psi}(x_2)\gamma^4\cancel{x}_2\cancel{A}}(x_2)\psi(x_2)\right]\\
\\
+N\left[\underleftrightarrow{\overline{\psi}(x_1)\gamma^4\cancel{x}_1\underleftrightarrow{\cancel{A}(x_1)\psi(x_1) \overline{\psi}(x_2)\gamma^4\cancel{x}_2\cancel{A}}(x_2)\psi}(x_2)\right]+
N\left[\overline{\psi}(x_1)\gamma^4\cancel{x}_1\underleftrightarrow{\cancel{A}(x_1)\underleftrightarrow{\psi(x_1) \overline{\psi}}(x_2)\gamma^4\cancel{x}_2\cancel{A}}(x_2)\psi(x_2)\right]
\\
\\
+N\left[\underleftrightarrow{\overline{\psi}(x_1)\gamma^4\cancel{x}_1\cancel{A}(x_1)\underleftrightarrow{\psi(x_1) \overline{\psi}}(x_2)\gamma^4\cancel{x}_2\cancel{A}(x_2)\psi}(x_2)\right]
\\
\\
+N\left[\underleftrightarrow{\overline{\psi}(x_1)\gamma^4\cancel{x}_1\underleftrightarrow{\cancel{A}(x_1)\underleftrightarrow{\psi(x_1) \overline{\psi}}(x_2)\gamma^4\cancel{x}_2\cancel{A}}(x_2)\psi}(x_2)\right]\, ,$
\\
\\
where $N[\cdots]$ is the normal order products and the symbol $\leftrightarrow$  under the two fields represents the vacuum expectation value of  the time order product of the two fields. The second and third terms  belong to spinor-vector (electron-photon) scattering (Compton scattering) and forth term  belongs to spinor-spinor scattering.
The spinor and vector fields expansion (\ref{psi expansion}) and (\ref{vector field expressed})
can be rewritten in the following forms:
\b \label{+,- part field}
\psi(x)=\psi^+(x)+\psi^-(x)
,\;\;
\overline{\psi}(x)=\overline{\psi}^+(x)+\overline{\psi}^-(x)
,\;\;
A(x)=A^+(x)+A^-(x)\;.
\e
Using the commutation and anti-commutation relations of creation and annihilation operators (\ref{anti-commutation relation}) and (\ref{commutation relation}), the act of $\psi^+ ,\overline{\psi}^-$ and $A^+$ on vacuum state are:
$$\psi^+|\Omega>=0,\; \;\; \overline{\psi}^-|\Omega>=0,\;\;\;  A^+|\Omega>=0.$$
In the following equation:
\b \label{4th term} N\left[\overline{\psi}(x_1)\gamma^4\cancel{x}_1\cancel{A}(x_1)\underleftrightarrow{\psi(x_1) \overline{\psi}}(x_2)\gamma^4\cancel{x}_2\cancel{A}(x_2)\psi(x_2)\right]
,\e
the symbole $\underleftrightarrow{\psi(x_1) \overline{\psi}}(x_2)$ is the spinor two-point function (\ref{spinor two-point function}):
$$\underleftrightarrow{\psi(x_1) \overline{\psi}}(x_2)\equiv\left<\Omega\left|\psi(x_1) \overline{\psi}(x_2)\right|\Omega\right>.$$ 
In relation (\ref{4th term}) the $\gamma , \;\cancel{x} $ and $\psi$ are in the matrix form, thus we write them in terms of their components. By inserting the (\ref{+,- part field}) in (\ref{4th term}), using transversality condition, $x.A=0$, and some simple calculation, one can see that the physical terms are as follows:
$$\mbox{physical\;terems}=\overline{\psi}^+_i(x_1)\cancel{A}^-_{jh}(x_1)\cancel{A}^+_{lr}(x_2) \psi^+_r(x_2)+\overline{\psi}^+_i(x_1)\cancel{A}^-_{lr}(x_2)\cancel{A}^+_{jh}(x_1) \psi^+_r(x_2)
$$
$$
+\overline{\psi}^-_i(x_1)\cancel{A}^-_{jh}(x_1)\cancel{A}^+_{lr}(x_2) \psi^-_r(x_2)+\overline{\psi}^-_i(x_1)\cancel{A}^-_{lr}(x_2)\cancel{A}^+_{jh}(x_1) \psi^-_r(x_2)\, .
$$
The two first terms will display the electron-photon (spinor-vector fields) scattering witch can be presented in figure \ref{f}:
\\
\\

By using the relation (\ref{spinor two-point function})
the $\cal S$ matrix for Compton scattering in ambient formalism of dS space in this approximation is obtained as:
$$ {\cal S}^{(2)}=-Q^2H^2 \int\int\ud\mu(x_1)\ud\mu(x_2)\gamma^4_{ij}(\cancel{x}_1S(x_1,x_2)\gamma^4\cancel{x}_2)_{hl}\;{\cal M}_{ijhl},$$
where $${\cal M}_{ijhl}\equiv \overline{\psi}^+_i(x_1)\cancel{A}^-_{jh}(x_1)\cancel{A}^+_{lr}(x_2) \psi^+_r(x_2)+\overline{\psi}^+_i(x_1)\cancel{A}^-_{lr}(x_2)\cancel{A}^+_{jh}(x_1) \psi^+_r(x_2)\, .$$
In this interaction the incoming state$|i>$ includes spinor and vector fields  and the outgoing state $|f>$ are another spinor and vector fields with different momentums and polarizations. Thus for incoming and outgoing states we have:
$$|i>=|1_{p,\sigma}^{a} , 1_{k,n}^{d}>=a^{\dagger}_\sigma(\xi_p)d^{\dagger}_n(\xi_k)|\Omega>,\;\; <f|=<1_{p',\sigma'}^{a} , 1_{k',n'}^{d}|=<\Omega|d_{n'}({\tilde{\xi}}'_k)a_{\sigma'}({\tilde{\xi}}'_p)\, ,$$
then the ${\cal S}^{(2)}$ matrix element on these states is:
$${\cal S}^{(2)}_{fi}=<f|{\cal S}^{(2)}|i>=-Q^2 \int\int\ud\mu(x_1)\ud\mu(x_2)\gamma^4_{ij}(\cancel{x}_1S(x_1,x_2)\gamma^4\cancel{x}_2)_{hl}\;<f|{\cal M}|i>,$$
where
$$<f|{\cal M}|i>={\cal N}_p{\cal N}_{p'}{\cal N}_k{\cal N}_{k'} \left[(Hx_1.{\xi}'_p)^{-2+i\nu} (Hx_2.\xi_p)^{-2-i\nu} \;\overline{\cal U}_i({\xi}'_p,\sigma'){\cal U}_r({\xi}_p,\sigma)\right]$$
$$
\times\left[ (Hx_1.{\xi}'_k)^{-2} (Hx_2.\xi_k)^{-1} \cancel{{\cal E}}_{1_{jh}}(x_1,{\xi}'_k,n')\cancel{{\cal E}}_{2_{lr}} (x_2,{\xi}_k,n)\right.$$
$$\left.\;\;\;\;\;\;\;\;\;\;\;\;\;\;\;\;\;+(Hx_1.\xi_k)^{-1} (Hx_2.\xi'_k)^{-2} \cancel{{\cal E}}_{1_{lr}}(x_2,{\xi}'_k,n')\cancel{{\cal E}}_{2_{jh}}(x_1,{\xi}_k,n)\right].$$ .

Hence one can write the scattering matrix element as:
$$
{\cal S}^{(2)}_{fi}=-{\cal N}_p{\cal N}_{p'}{\cal N}_k{\cal N}_{k'}Q^2 \int\int\ud\mu(x_1)\ud\mu(x_2)\;\;\overline{\cal U}_i(\xi'_p,\sigma')\;\gamma^4_{ij}\;(\cancel{x}_1S(x_1,x_2)\gamma^4\cancel{x}_2)_{hl}$$
$$\times\left [(Hx_1.{\xi}'_p)^{-2+i\nu} \;(Hx_2.\xi_p)^{-2-i\nu}\right] $$
$$\times \left[(Hx_1.{\xi}'_k)^{-2} (Hx_2.\xi_k)^{-1} \cancel{{\cal E}}_{1_{jh}}(x_1,{\xi}'_k,n')\cancel{{\cal E}}_{2_{lr}} (x_2,{\xi}_k,n)\right.+$$
\b\label{scattering matrix}
\left. (Hx_1.\xi_k)^{-1} (Hx_2.\xi'_k)^{-2} \cancel{{\cal E}}_{1_{lr}}(x_2,{\xi}'_k,n')\cancel{{\cal E}}_{2_{jh}}(x_1,{\xi}_k,n)\right]
 {\cal U}_r(\xi_p,\sigma) ,\e
where ${\cal N}_p\;,\;{\cal N}_{p'}\;,\;{\cal N}_k\;,\;{\cal N}_{k'}$ are normalization constant of incoming and outgoing spinor and vector fields \cite{bagamota,gagarota,gata}.

The integral (\ref{scattering matrix}) is taken over the dS space-time and all elements are known, it can be calculated by using the Bros-Fourier transformation. In Minkowski flat space, the Fourier transforms of most functions are known, but in dS space, apart from the scalar field which has been partly done by Bros et al. \cite{brmo,brmo03}, not much work has been done on this transformation. Therefore, using numerical methods, one may find a solution for the integral that determines the curvature effects in QED.  It will be investigated in the subsequent articles.

\section{Flat limit} \label{mili}
The dS manifold is a four dimensional hyperboloid with radius $H^{-1}$ and for larger radius the curvature of space-time decreases, so when $H\rightarrow0$ the curvature of space-time disappears and space-time matches with the flat Minkowski space-time.
We can parametrize the $\xi_p$ in terms of the massive particle four-momentum $p^\mu$, which is suitable for the null curvature limit:
$$\xi_p=\left(\xi^{0} , \overrightarrow{\xi} , \xi^{4}\right)=\left(\dfrac{p^{0}}{m}=\sqrt{\dfrac{\overrightarrow{p}^2}{m^2}+1} , \dfrac{\overrightarrow{p}}{m} , -1\right)\,.$$
We can also express the dS point $x^\alpha$ in terms of the intrinsic coordinates $X^\mu=\left(X^{0}=ct\; , \overrightarrow{X}\right)$ as:
$$x^\alpha=\left(x^0=H^{-1}\sinh(HX^0)\; , \overrightarrow{x}=H^{-1}\dfrac{\overrightarrow{X}}{\lVert\overrightarrow{X}\lVert}\cosh(HX^0)\sinh(H\lVert\overrightarrow{X}\lVert)\; , x^4=H^{-1}\cosh(HX^0)\cosh(H\lVert\overrightarrow{X}\lVert)\right),$$
where ${\lVert\overrightarrow{X}\lVert}=(X_1^2+X_2^2+X_3^2)^{\frac{1}{2}}$. Therefor, it is straightforward to show that \cite{ta97}:
$$\lim_{H\rightarrow 0}(x\cdot\xi_p)^{-2-i\nu}=e^{-ip.X} ,\;\; \lim_{H\rightarrow 0}(x\cdot\xi_p)^{-2+i\nu}=e^{ip.X},$$
$$\lim_{H\rightarrow 0}H\cancel{x}=\lim_{H\rightarrow 0}H\eta_{\alpha\beta}\gamma^\alpha x^\beta=-\gamma^4,\;\;\;\lim_{H\rightarrow 0}\cancel{\xi}_p=\gamma^\mu p_\mu+\gamma^4, $$
where $\mu=0,1,2,3 $ and $\alpha=\beta=0,1,2,3,4$. For the dS vector field we can parametrize the $\xi$ in terms of the massless particle four-momentum $k^\mu$ as:
$$ \xi_k^\alpha = \left(k^0,  \vec{k} , H \right) ,\;\; k^0 > 0,\; \;\;\; (k^0)^2- \vec{k}\cdot \vec{k}=H^2 .$$
For dS massless vector field one can showthat the polarization states are \cite{gagarota,ta97}:
$$\lim_{H\rightarrow 0}{\cal E}_{2\alpha}(x,\xi,n)=\epsilon_\mu^{(n)},\;\;\;  \lim_{H\rightarrow 0}{\cal E}_{1\alpha}(x,\xi,n)=\epsilon_\mu^{*(n)}\, ,$$
where
$$\epsilon^{(n)}.k=0,\;\;\;\epsilon^{(n)}.\epsilon^{*(n')}=\eta^{nn'},\;\;\;\sum_{n=0}^{3}\dfrac{\epsilon_\mu^{(n)}\epsilon_\nu^{*(n)}}{\epsilon^{(n)}.\epsilon^{*(n)}}=\eta_{\mu\nu}\,.$$
Two-point functions of spinor and vector fields at $H\rightarrow 0$ limit are matched with Minkowski counterparts \cite{ta97,ta96,bagamota,gagarota}. Using the definition $\overline{{\cal U}}={\cal U}^\dagger\gamma^0\gamma^4$ and (\ref{minkofski gamma}), the null curvature limit of scattering matrix (\ref{scattering matrix}) is:
$$\lim_{H\rightarrow 0}{\cal S}^{(2)}_{fi}={\cal N}_p{\cal N}_{p'}{\cal N}_k{\cal N}_{k'}Q^2\dfrac{1}{(2\pi)^4}\int\ud^4x_1\int\ud^4 x_2\int\ud^4q$$
$$\times \{e^{-ik'.x_1}e^{ik.x_2}e^{iq.(x_2-x_1)} e^{-ip'.x_1}e^{ip.x_2}\;\;\overline{\cal U}_i^{\prime}(p',\sigma')\;\gamma_{ij}^{4}\cancel{\epsilon}^{*(n')}_{jh}(k')\;\gamma_{ht}^4\;\dfrac{\left(q^\mu\gamma'_{\mu}+m\right)_{tl}}{q^2-m^2+i\epsilon}\cancel{\epsilon}^{(n)}_{lr}(k){\cal U}_r(p,\sigma)$$
$$+e^{ik.x_1}e^{-ik'.x_2}e^{iq.(x_2-x_1)} e^{-ip'.x_1}e^{ip.x_2}\;\;\overline{\cal U}_i^{\prime}(p',\sigma')\;\gamma_{ij}^{4}\cancel{\epsilon}^{(n)}_{jh}(k)\;\gamma_{ht}^4\;\dfrac{\left(q^\mu\gamma'_{\mu}+m\right)_{tl}}{q^2-m^2+i\epsilon}\cancel{\epsilon}^{*(n')}_{lr}(k'){\cal U}_r(p,\sigma)\}.$$
After a straightforward calculation, one obtained:
$$\lim_{H\rightarrow 0}{\cal S}^{(2)}_{fi}={\cal N}_p{\cal N}_{p'}{\cal N}_k{\cal N}_{k'} (iQ)^2 
(2\pi)^4\;\;\delta(k+p-k'-p')$$
$$\times \{\;{\cal U}_i^{\dagger(\sigma')}(p')\;\gamma'^0_{ij}\;\epsilon^{*(n')}_{\tau}(k')\gamma'^\tau_{jh}\;\dfrac{\left(k^\mu\gamma'_{\mu}+p^\mu\gamma'_{\mu}+m\right)_{hl}}{(k+p)^2-m^2+i\epsilon} \;\epsilon^{(n)}_{\nu}(k)\gamma'^\nu_{lr}\;{\cal U}_r^{(\sigma)}(p)$$
$$+\;{\cal U}_i^{\dagger(\sigma')}(p')\;\gamma'^0_{ij}\;\epsilon^{(n)}_{\tau}(k)\gamma'^\tau_{jh}\;\dfrac{\left(k^\mu\gamma'_{\mu}+p^\mu\gamma'_{\mu}+m\right)_{hl}}{(k+p)^2-m^2+i\epsilon} \;\epsilon^{*(n')}_{\nu}(k')\gamma'^\nu_{lr}\;{\cal U}_r^{(\sigma)}(p)\;\}\, ,$$
which is exactly the Minkowski QED results.

\section{Conclusion} \label{conclu} 
The effect of gravitational field or curvature on the interaction of electron-photon in dS ambient formalism is studied. The scattering matrix explicitly includes the sentences that show the curvature effects. Since the Bros-Fourier transform in dS space-time has not been defined explicitly, the integral (\ref{scattering matrix}) may not be calculated analytically for obtaining the scattering matrix. This scattering matrix should be calculated by numerical methods, which will be investigated in the next articles. In the null curvature limit ($H\rightarrow 0$) the Minkowski results are obtained. These calculations can be the starting point for considering the interactions of the quantum electron-photon fields with the classical gravitational field in dS ambient space formalism, which may be examining with the experimental test.

\vspace{0.5cm} 

{\bf{Acknowledgments}}:  The authors wish to express his particular thanks to  M. Amiri, M. Rastiveis,  R. Raziani and S. Tehrani-Nasab for discussions.

\end{document}